\documentclass[12pt,preprint]{aastex}

\def\au{{\rm AU}} 
\def\kms{{\rm km}\,{\rm s}^{-1}}

\def\max{{\rm max}}
\def\peri{{\rm peri}}

\begin{document}
\title{Gravitational Pulse Astronomy}

\author{Andrew Gould}
\affil{Department of Astronomy, Ohio State University,
140 W.\ 18th Ave., Columbus, OH 43210, USA; 
gould@astronomy.ohio-state.edu}

\begin{abstract}

Thompson has argued that the Kozai mechanism is primarily responsible
for driving white-dwarf binary mergers and so generating type Ia
supernovae (SNe).  If so, the gravitational wave signal from these
systems will be characterized by isolated repeating pulses that are
well approximated by parabolic encounters.  I show that it is
impossible to detect these with searches based on standard assumptions
of circular binaries, nor could they be detected by analogs of the
repeating-pulse searches that have been carried out at higher
frequencies, even if these were modified to barycentric time as a
function of putative sky position.  Rather, new search algorithms are
required that take account of the intrinsic 3-body motion of the
source as well as the motion of the Earth.  If these eccentric
binaries account for even a modest fraction of the observed SN rate,
then there should be of order 1 pulse every 20 seconds coming from
within 1 kpc, and there should be of order 200 detectable sources in
this same volume.  I outline methods of identifying these sources both
to remove this very pernicious background to other signals, and to
find candidate SN Ia progenitors, and I sketch practical methods to
find optical counterparts to these sources and so measure their masses
and distances.

\end{abstract}


\section{{Introduction}
\label{sec:intro}}

Except near final in-spiral, binary-star gravitational-wave (GW) sources
are strictly periodic, and therefore 
a Fourier transform of the amplitude can be used to detect these
sources.  The only nuance is that one must also search over the
sky by converting to a barycentric frame as a function of sky position.
And, once a source is located, one can then ``tune up'' the signal
by fitting for the phase and spatial orientation of the binary orbit.

Until very recently, it was believed that the vast majority of
individually identifiable binary sources would be on circular orbits.
In this case the signal is sinusoidal, so a Fourier transform is not
merely convenient, it also contains a matched filter to the
signal, and so is ``optimal''.  However, \citet{thompson10} has 
shown that the great majority of white-dwarf (WD) binary
sources probably have highly eccentric orbits.  Within a
``Fourier framework'', the resulting signal 
is represented by a discrete power spectrum, which
covers the range $[2\pi/P,\Omega]$, where $P$ is the orbital period,
$(\Omega P/2\pi)= (1-e)^{-3/2}$ and $e$ is the eccentricity.  

However, for eccentric orbits, Fourier transform no longer
provides anything like a matched filter.  For isolated binaries,
one could in principle simply extend the search by including
eccentricity and time of periastron as additional parameters.  Then,
since such systems are still strictly periodic (or almost so),
one could fold the, say, 5-year GW data stream by
the period (modulated by the annual motion of 
Earth as a function of putative sky position) to optimize signal
detection.

Unfortunately, the eccentric WD binaries predicted by
\citet{thompson10} are, by their very nature, not isolated.  The high
eccentricities are induced by the presence of a third body with a
separation that is $\sim$ 10--100 times larger than the semi-major
axis of the eccentric binary.  The lower limit of this range is
required for hierarchical stability and the upper limit to make the
Kozai mechanism effective.  Thus, several extensions of existing
search techniques will be required to detect these objects.  Note
that the low mass density of WDs implies low-frequency waves, which
are only detectable from space.

Some intuition into this problem is gained by noting that the signal
is completely dominated by pulses near periastron, and that during
these passages the eccentric orbit is hardly distinguishable from a
parabolic one.  Here I develop this approximation and apply it to the
problem to extracting science from gravitational pulse signals.

\section{{The Parabolic Pulse}
\label{sec:parabolic_pulse}}

Parabolic orbits may be parametrized by
\begin{equation}
{r\over b}= 1 + {\psi^2\over 2},
\quad
{x\over b}= 1 - {\psi^2\over 2},
\quad
{y\over b} = \sqrt{2}\psi,
\quad
\Omega t = \psi + {\psi^3\over 6}.
\label{kepler}
\end{equation}
Here, ${\bf r}\equiv (x,y)$ is the separation between the two masses,
$b$ is the distance at periastron, $\psi$ is a parameter,
$\Omega=\sqrt{GM/b^3}$, $M$ is the total mass, and $t$ is time.  In the
quadrupole approximation, the gravitational strain tensor $h_{ij}$ is given
by (\citealt{mtw}, MTW, Eqs.~36.3, 36.47)
\begin{equation}
h_{ij} = {2\over D}\, {d^2 I^{\rm traceless}_{ij}\over dt^2}
\qquad
I^{\rm traceless}_{ij}\equiv I_{ij} - {1\over 3}\delta_{ij}\sum_k I_{kk},
\qquad
(G=c=1)
\label{qdef}
\end{equation}
where $I^{\rm traceless}_{ij}$ is the traceless part of the moment of inertia
tensor $I_{ij}$, and $D$ is the source distance.

Differentiation yields
\begin{equation}
{h_{xx}\over A} = {3y^2 r - 6 x^2 b - 2 r^2 b\over 3 r^3},
\qquad
{h_{yy}\over A} = {12 x b^2 - 2 r^2 b\over 3 r^3},
\qquad
{h_{zz}\over A} = -{2 b\over 3 r},
\qquad
{h_{xy}\over A} = {y^3\over r^3},
\label{eqn:heval}
\end{equation}
where
\begin{equation}
A\equiv 2{M_1 M_2\over b D}
= 2.1\times 10^{-21}
{M_1 M_2\over M_\odot^2}\,\biggl({b\over 0.1\,R_\odot}\biggr)^{-1}
\biggl({D\over {\rm kpc}}\biggr)^{-1}
\label{eqn:adef}
\end{equation}
is the normalization of the GW amplitude, $M_1$ and $M_2$
are the two masses, and $M=M_1+M_2$.  These waveforms are shown
in Figure~\ref{fig:waveform}.  As expected, they are strongly concentrated
is a small interval of time $\pm\Omega^{-1}$ near periastron, where
\begin{equation}
\Omega^{-1} = 
\sqrt{b^3\over M} =
50\,{\rm s}\biggl({b\over 0.1\,R_\odot}\biggr)^{3/2}
\biggl({M\over M_\odot}\biggr)^{-1/2}.
\label{eqn:omegadef}
\end{equation} 

The total energy of the pulse is most easily calculated by
taking the limit $e\rightarrow 1$ in the standard formula for the
mean luminosity of an eccentric binary (MTW, Eq.~36.16a,b)
\begin{equation}
E_{\rm pulse} = {85\pi\over 16}M_1^2 M_2^2\sqrt{M\over 2b^7}.
\label{eqn:epulse}
\end{equation}
This energy may be directly compared to the potential energy at periastron,
\begin{equation}
Q\equiv 
{M_1 M_2/b\over E_{\rm pulse}} = 4\times 10^{10}
\biggl({M_1\over M_\odot}\biggr)^{-1}
\biggl({M_2\over M_\odot}\biggr)^{-1}
\biggl({M\over M_\odot}\biggr)^{-1/2}
\biggl({b\over 0.1\,R_\odot}\biggr)^{5/2}.
\label{eqn:qdef}
\end{equation}

 Hence, if the \citet{thompson10} Kozai mechanism is responsible for even
one Milky Way supernova (SN) Ia per 500 years, then there must be one
pulse per $(500 {\rm yr})/Q \sim 0.2$ seconds.  And even supposing
that only 1\% of these are within 1 kpc of the Sun, there would
still be one such ``loud'' pulse every 20 seconds.  Hence, disentangling
these pulses is important both from the standpoint of understanding
the sources and removing them as a background.

In addition, if this mechanism is indeed important for generating
SNe Ia, then it will almost certainly create a large population of
in-spiraling WDs that lack sufficient combined mass to explode
but still contribute to the gravitational pulse cacophony.

\section{{Pulse Search Strategy I: Isolated Eccentric Binaries}
\label{sec:strategy}}

I begin by analyzing the simplified case of isolated eccentric binaries.
As just noted, the real sources are not isolated,
but this permits me to explore most of the relevant physics
before introducing this additional complication.

Then because the pulses are periodic, they would in principle turn up
in a Fourier-transform type search.\footnote{Such searches must be
conducted separately for every independent direction on the sky by
transforming to the barycentric time for that particular direction.  I
will discuss searching in many directions
below, but for the moment will assume one direction has been chosen
and that time has been corrected to barycentric.}  However, the
signal-to-noise ratio (SNR) of such a search would be degraded
relative to a matched-filter search by roughly
$(P\Omega/2\pi)^{1/2}\sim (1-e)^{-3/4}$.  For example, for $e=0.99$,
the degradation would be a factor $\sim 30$, meaning that roughly 30
times smaller signals would be accessible to a matched-filter search
than a Fourier-transform search.

Of course, there is some cost associated with enhanced sensitivity.
For each period $P$, one would need to consider $P\Omega$
different independent phases, where I adopt $\Omega^{-1} =40\,$s as the
typical value.  And one
would have to step through the periods at 
$\Delta P= P/(\Omega T_{\rm mission})$, leading to a total 
(for each sky position) of 
$N_{\rm try} =T_{\rm mission}P_{\rm max}\Omega^2= 3\times 10^{10}$ independent
trials, where I adopt $T_{\rm mission}=5\,$yr as the lifetime of the
mission and $P_{\rm max}=3\,$days as the maximum period that will be
searched.  And for each such trial, one would have to try
several, perhaps 10, independent pulse widths $\Omega^{-1}$.  
Efficient algorithms have already been developed for carrying out
such searches in studies of both periodic GW sources
(\citealt{abadie10} and references therein) and transiting planets 
\citep{kovacs02}.  However, 
there is also a cost associated with the added risk of false positives.
If one assumes Gaussian
noise, then the minimum SNR required to avoid false positives is $\sim
\sqrt{2\ln N_{\rm try}}$.
If, for example, the number of independent trials goes from 
$10^4$ to $10^{15}$, then
the minimum SNR goes roughly from 4.3 to 8.3, i.e., a factor of 2.
This is very modest compared to the factor $\sim 30$ gain in
sensitivity due to matched filters.

\section{{A Rough Estimate of LISA Source Counts}
\label{sec:lisa}}

To proceed further, I consider the specific example of the
Laser Interferometer Space Antenna (LISA, \citealt{prince09}), 
a proposed
space mission composed of 3 antennae separated by $L=0.033\,\au$
in an Earth-like heliocentric orbit.  Lisa sensitivity peaks
for sinusoidal periods at $P\sim 150$ seconds at a threshold
of $h\sim 10^{-23}$, with a full-width half-maximum (FWHM) of about 
1 decade.  Hence, it is extremely well-matched to pairs of $0.7\,M_\odot$
WDs
with periastra $b\sim 0.1\,R_\odot$, which have pulse widths
corresponding to ``sinusoidal periods'' of
about $2\pi/\Omega\sim 260$ seconds.  That is,
for WD binaries with $M=1.4\,M_\odot$, the
FWHM of LISA sensitivity corresponds to
\begin{equation}
b = 0.07^{+0.08}_{-0.04}\,R_\odot \qquad \rm (FWHM).
\label{eqn:binterval}
\end{equation}
However, when comparing forecasted LISA sensitivity to 
Equation~(\ref{eqn:adef}), one must adjust by a factor 
$\sim (1-e)^{-3/4}$, to account for the fact that the signal
only builds up once per period $P$, not once per $2\pi/\Omega$.
Even so, for $e=0.99$, LISA is sensitive to strains just 30 times
higher than its nominal sensitivity, i.e., $h\sim 3\times 10^{-22}$.  
Comparing to Equation~(\ref{eqn:adef}),
and noting from Figure~\ref{fig:waveform} that the signal
peaks at a few times $A$,  it is clear that LISA will be sensitive
to such binaries to several kpc.  

The period distribution of sources is governed by a complex interplay
of pre-WD and post-WD Kozai, binary evolution, GW emission,
and tidal effects \citep{thompson10}, which are beyond the scope of
this Letter.  However, one can make a simple estimate
by assuming that all pulses are the same.  Then energy loss is governed
by $dE/dt\propto 1/P$, which implies that the cumulative number of systems
is $N(P)\propto P^{1/3}$.  Hence, the {\it population} is weakly dominated
by high-period systems, while pulse generation is dominated by tight 
systems: $(dN/d\ln P)/P \propto P^{-2/3}$.  This is important.  It means
that most of the ``noise'' is concentrated in the highest SNR objects,
so easiest to detect.
But it also means that the most interesting sources are in the
long-period tail.  If we assume a typical ``injection eccentricity''
$e_\max = 0.99$, then 2/3 of the systems will have $e\ga 0.9$.
And if the ensemble of sources is responsible for 1 SNe Ia per 50,000 years
within 1 kpc, then there are of order 200 such binaries within 1 kpc.
Obviously, this is a very crude estimate, but the point is that
there are potentially many more sources than there would be
to supply the same SN Ia rate from circular orbits.

\section{{Signal-to-Noise Ratio Estimates}
\label{sec:SNR}}

\begin{figure}
\plotone{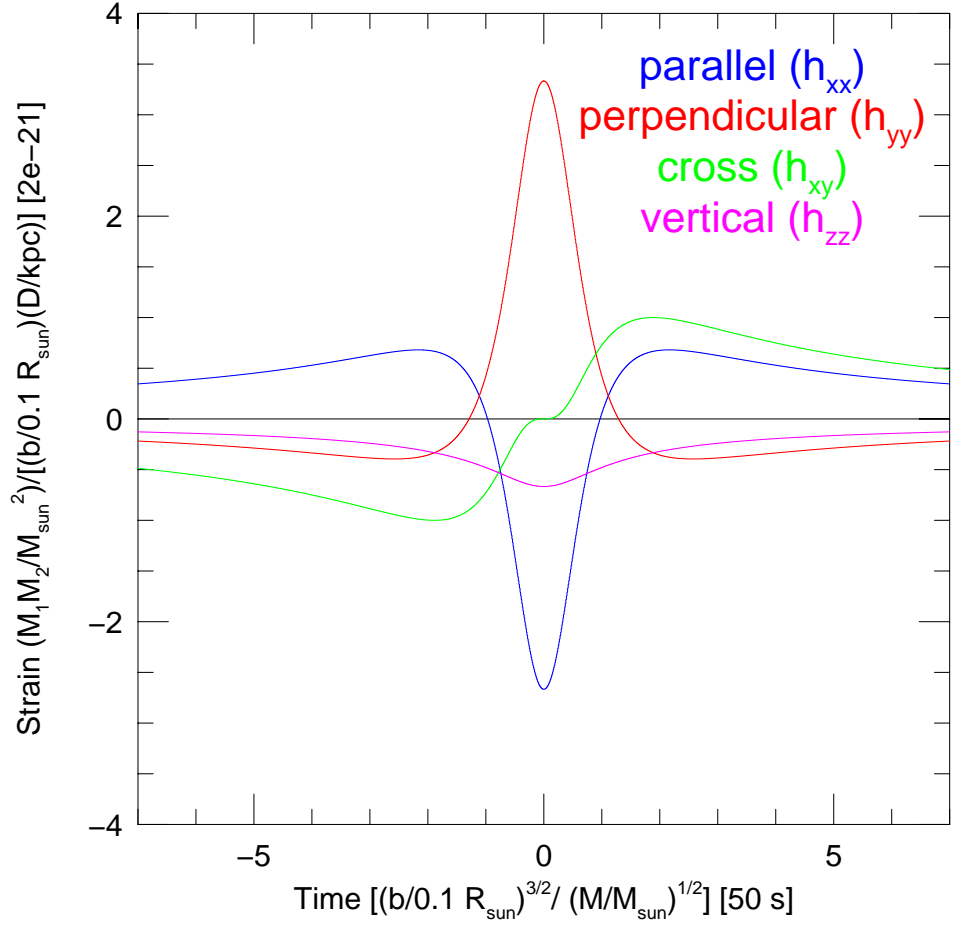}
\caption{\label{fig:waveform}
Four non-vanishing components of the gravitational waveform due to
a parabolic encounter between two masses, which is the limiting
case of a highly elliptical orbit.  The periastron is on the $x$
(``parallel'') axis and orbital plane is determined by this and
the ``perpendicular'' axis.  These and the ``vertical'' components
sum to zero at all times, since the matrix is traceless.  To find the
waveform seen at Earth, one would have to rotate according to the
Euler angles and then project out only the transverse components.
The wave-amplitude $A$ and frequency $\Omega$ are normalized
to units of $M_\odot$, $0.1\,R_\odot$ and kpc, for masses, periastra,
and distance.  
}
\end{figure}

To make a more precise estimate of the expected SNR, I integrate the
square of the profiles shown in Figure~\ref{fig:waveform}, averaging
over all Euler orientations, 
and restricting the integral to $\pm 3\Omega^{-1}$.
I assume that the Lisa sweet spot permits detection of sources with
rms strain $h=A_0=1.0\times 10^{-23}$ in $T_0=1\,$yr of observation
at $5\,\sigma$ (\citealt{prince09}, pp 2, 21).
I then find a SNR {\it for a single pulse} of
\begin{equation}
{\rm SNR}_{\rm pulse} = 20{A\over A_0}(\Omega T_0)^{-1/2}
\simeq 5.4\,{M_1 M_2\over M^{1/4}M_\odot^{7/4}}
\biggl({b\over 0.1\,R_\odot}\biggr)^{-1/4}
\biggl({D\over {\rm kpc}}\biggr)^{-1} F(\Omega)
\label{eqn:snrpulse}
\end{equation}
where $F(\Omega)$ is the functional form of LISA sensitivity \citep{prince09}
normalized to unity at maximum sensitivity at $\Omega^{-1}=24\,$s.
Note that this formula basically scales  $\propto M_1 M_2/D$.
The integrated SNR from summing all the pulses observed during the
mission is larger by
\begin{equation}
{\rm SNR} = {\rm SNR}_{\rm pulse} \sqrt{T_{\rm mission}\over P}.
\label{eqn:snrpulse2}
\end{equation}
For example, for $\Omega^{-1} = 40\,$s, and $e=0.99$, we have
$P= 2\pi\Omega^{-1}(1-e)^{-3/2}= 3\,$days, so 
${\rm SNR} = 25\,{\rm SNR}_{\rm pulse}$.  Thus, even requiring a
$9\,\sigma$ detection (see below) would permit detections throughout
the Galaxy, provided that multiple pulses from the same source could
be so identified and thus ``co-added''.  Moreover,
individual pulses from within $D\la 1\,$kpc would be detectable.
They, and their weaker cousins from farther away would 
constitute an incredible data mine if they could be interpreted, but
a vast cacophony of noise if they could not.  Since the great majority
of this ``noise'' is due to the small number of binaries that have
already been driven to shorter periods, it is important to note
that for $e<0.1$, ${\rm SNR} > 140 {\rm SNR}_{\rm pulse}$, implying
that these sources are much easier to extract and ``remove''.

Integrating the square of the time derivative of the wave forms in
Figure~\ref{fig:waveform} yields the timing precision of each pulse,
$\sigma_t \simeq 0.8\,\Omega^{-1}{\rm SNR}_{\rm pulse}^{-1}$.
Hence, if all the $N_{\rm pulse}=T_{\rm mission}/P$
pulses from a single source can be successfully aligned,
the angular direction precision from fitting
to the correct barycentric pulse-delay pattern is (in radians)
\begin{equation}
\sigma_\theta = \sqrt{2\over N_{\rm pulse}}{\sigma_t\over \au}
= {0.09\over {\rm SNR}}\, (40\,{\rm s}\,\Omega)^{-1}.
\label{eqn:sigmatheta}
\end{equation}
That is, sources within $D\la 1\,$kpc could be located within a few
arcmin.  Position measurements from LISA orbital motion alone would
suffer an exact degeneracy in ecliptic latitude.  The directional
information relative to the spatial orientation of LISA is about
$\au/L=30$ times worse than the orbital information, but it does not
suffer from this ecliptic degeneracy.  It would therefore be sufficient to
break the orbit-based-direction degeneracy for nearer sources, but not
more distant ones.

\section{{Pulse Search Strategy II: Triples}
\label{sec:strategy2}}

As mentioned in the Introduction, the Thompson-Kozai WD
binaries are all embedded in triples with semimajor-axis
ratios $a_2/a_1\sim$ 10--100.  Hence $0.3\la a_2/\au\la 10$.
To illustrate the problems posed by this, I consider the case
$M_1=M_2=M_3=0.7\,M_\odot$, where $M_3$ is the third body.  
Then, the binary will orbit the
center of mass with an amplitude $a_2/3$, which means that, if not
corrected, the pulse signal will drift by 
$\sim 170\,{\rm s} (a_2/\au)\sin i$, where $i$ is the inclination.
This is many times larger than the pulse width, especially considering
that for $a_2<6\,\au$, the system will complete at least half an orbit
in 5 years.  Hence, if the orbit around the third body is not included,
the pulses will not align, even approximately, and the signal cannot
be recovered.

Hence, for each sky position (and its corresponding barycentric
correction), and each combination of $(\Omega,P,t_\peri)$, where 
$t_\peri$ is the time of pericenter of the inner binary, one must
conduct at least a 3-parameter search, corresponding to the
period, phase, and amplitude of a sinusoidal orbit about $M_3$.
Then the number of such trials would be
$\sim [(a_{2,\rm max}/3)\Omega]^3\sim 10^4$, where 
$a_{2,\rm max}\sim 6\,\au$.  For $a_2>a_{2,\rm max}$, a simpler
one parameter uniform-acceleration model would be adequate.

Thus, the full search would be over $4\pi\,(\au\,\Omega)^2=2\times
10^3$ sky positions, $P_{\rm max} T_{\rm mission}\Omega^2\sim 3\times
10^{10}$ inner binary periods and phases, $[(a_{2,\rm
max}/3)\Omega]^3\sim 10^4$ outer binary trials, and perhaps 10
different pulse widths.  This is $10^{19}$ independent trials, which
requires a detection threshold of SNR $>9$.  The sheer volume of
computations is forbidding.  As already mentioned, there are efficient
algorithms for doing the $3\times 10^{10}$ period search calculations.
The problem is that these must each be done for $2\times 10^8$
different configurations.  Here I will simply assume that Moore's Law
will handle this problem, although there is probably room for
algorithmic improvements as well.  In particular, the shorter-period
binaries that dominate the ``noise'' will be detectable from
subintervals $T\ll T_{\rm mission}$, which will permit elimination of
the outer-binary trials, and even allow standard Fourier techniques in
many cases.  Removal of this dominant ``noise'' will be essential to
finding the more numerous high-eccentricity binaries.

\section{{Extracting Science}
\label{sec:science}}

What parameters can be extracted from such observations, and how
can the remaining degeneracies be resolved?   I will argue that
full resolution requires identification of optical counterparts.
The counterparts that are easiest to identify are 3-WD
systems, and these are also the most likely to yield key
spectroscopic data leading to complete resolution.  I therefore
focus first on these systems.  There will of course be a huge number
of systems without counterparts, which could be subjected to
statistical analysis, but the analysis of that problem
is beyond the scope of this Letter.

First, I review the observables. The best-fit barycentric
correction gives the direction on the sky, and the waveform gives
the three Euler angles of the binary (up to discrete degeneracies
due to the quadrupole nature of GWs).  These will be of use
further below but I ignore them for the moment.  There are also
(at least) three orbital parameters for third body.  The remaining
four parameters that can be measured are $A$, $\Omega$, $P$, and
$t_\peri$.  The period is a parameter of
fundamental interest since it helps classify the systems according
to their evolutionary state and rate of progression toward WD
collisions.  Together with $\Omega$, the period also gives the
eccentricity, which is again of independent interest and also
enables a more precise waveform calculation (although I expect that this
will be a very minor consideration for high eccentricity orbits).
The time of periastron will also be important further below, but
will be ignored for the moment.  This leaves two observables
\begin{equation}
A= 2{M_1 M_2\over b D},
\qquad
\Omega^2 = {M_1 + M_2\over b^3}
\label{eqn:Aomega}
\end{equation}
that combine four physical parameters, $M_1$, $M_2$, $b$, and $D$.
It is important to keep in mind $\Omega$, $P$, and $t_\peri$ are
all robustly determined from distinct features in the data, while $A$ is 
somewhat degenerate with the Euler angles, particularly the inclination
and longitude of nodes.  However, I ignore this degeneracy for the moment
and assume that $A$ is also well determined.  Then, since there
are four parameters and two measurements, there remain two degeneracies 
to be broken for which two additional pieces of information are required.  
Real additional ``information'' can only be derived from counterparts,
but at the outset, one can gain
a rough idea of $b$ and $D$ simply by assuming, e.g., 
$M_1=M_2=0.6\,M_\odot$. Of course, such an assumption would 
make it impossible to derive the masses and thus determine
whether the system was a viable SN Ia progenitor, which is
arguably the most interesting science potential of the sample.
But it would give a rough estimate of the distance.

To make any further progress would require a catalog of WD
candidates over the estimated distance range of the sample.
Optical identification of WDs would be extremely difficult
if the third star in the system were a main-sequence star.  
However, a substantial
fraction of tertiaries are likely to be descendants of intermediate
mass stars, and so themselves WDs.
The most efficient way to construct a catalog of WDs
(or multiple WDs) is a reduced
proper motion survey (e.g., \citealt{salim03}).  
This will be a natural by-product of the
Large Synoptic Survey Telescope (LSST) for about 3/4 of the sky.
Since WDs are typically $M_V\la 15.5$, LSST should reach
several hundred pc or more, depending on its exact performance.
For example, \citet{lsst} expect that at $r=24$ (so $D\sim 400\,$pc,
allowing for extinction $A_r\sim 0.5$) the ``main'' LSST survey (1/2 sky)
will achieve a proper motion
precision of $1\,\rm mas\,yr^{-1}$, corresponding to a transverse
velocity error of $2\,\rm km\,s^{-1}$.  This is about 5 times
better than is required to reliably identify a WD on a reduced
proper motion diagram.  Hence, even in the less-well covered northern
1/4 of the sky, LSST reduced proper motion diagrams should be
adequate to about 400 pc.

Recall from Section~\ref{sec:SNR} at these distances SNR $\ga 60$ and
hence the position is known to $\la 5'$.  Since the surface
density of WDs to this distance is only $\sim 25\,\rm deg^{-2}$,
there would be only a few candidate objects consistent with
the position determined from GWs, even allowing for
a factor $\sim 2$ uncertainty in the reduced-proper-motion distance
estimates (and the smaller error in $A$ for these high SNR objects).

A spectrum taken near periastron would, by itself, positively
identify the system as the LISA counterpart because the velocity
difference between the two components would be of order $1000\,\kms$.
Even, if the system turned out to be single-lined, the observed
(brighter) component would be the less massive, and therefore would
be moving at extremely high velocity.  If the system were double lined,
the radial velocity curve would give the mass ratio
$M_1/M_2$, while the spectra themselves would yield individual masses.
The ratio of these could be checked against the radial-velocity value,
while the sum would yield $\sin i$, which could be checked against
the value determined from the GW pulse profile.
A similar exercise could be applied to the third WD,
yielding a comprehensive picture of the entire system.
Even if the more massive (so fainter) WD were beyond the
detection limit, the degeneracies could be partly resolved by
obtaining a trigonometric parallax.

These determinations can probably be made for a large fraction
of WD triples out to 400 pc, which plausibly number in the
dozens.  Moreover, since their individual
pulses can be detected, the algorithms
to identify these sources are much simpler than those outlined
above.  

Finally, I note that WD binaries with main-sequence companions
can also be positively identified from the correspondence between
the pulse timing residuals and the radial velocity curve of the
companion.  While the WDs would not be directly observable,
precise astrometric measurements combined with a spectroscopic
mass estimate of the companion would yield both $D$ and $M_1+M_2$,
which (with Eq.~[\ref{eqn:Aomega}]) permit a complete solution.

\acknowledgments

I thank Todd Thompson, Rubab Khan, and Scott Gaudi for useful discussions.  
The manuscript benefited greatly from careful review by the referee.
Work supported by NSF AST-0757888

\end{document}